\documentclass[
nofootinbib,
amsmath,amssymb,
 aps,
prd,
superscriptaddress,
10pt,
]{revtex4-2}
\usepackage{tikz}
\usepackage{graphicx}
\usepackage{comment}
\usepackage{footmisc}
\usepackage{amsmath}
\usepackage{dcolumn}
\usepackage{bm}
\usepackage{appendix}
\usepackage{physics}

\begin{document}
\preprint{APS/123-QED}
  \title{Gertsenshtein effect on the spacetime curved by background magnetic field with geometric optics}
\newcommand{\TokyoAffil}{Department of Physics, Institute of Science Tokyo, 2-12-1 Ookayama, Meguro-ku, Tokyo 152-8551, Japan}
\newcommand{\UtahAffil}{Department of Physics and Astronomy, University of Utah, 270 S 1400 East \#E2108, Salt Lake City, UT 84112, USA}

\author{Ryutaro Tomomatsu}
\email{tomomatsu.r.b02d@m.isct.ac.jp}
\affiliation{\TokyoAffil}

\author{Teruaki Suyama}
\email{suyama@phys.sci.isct.ac.jp }
\affiliation{\TokyoAffil}

\author{Paolo Gondolo}
\email{paolo.gondolo@utah.edu}
\affiliation{\UtahAffil}
\affiliation{\TokyoAffil}

\begin{abstract}
When electromagnetic (or gravitational) waves propagate in the presence of a background magnetic field, a portion of the waves converts into gravitational (or electromagnetic) waves. This phenomenon, known as the (inverse) Gertsenshtein effect, is typically analyzed in Minkowski spacetime, neglecting the spacetime curvature induced by the magnetic field itself.
This paper investigates, for the first time, the influence of spacetime curvature on the (inverse) Gertsenshtein effect. To this end, we first determine the metric perturbation from Minkowski spacetime up to second order in the magnetic field strength, assuming cylindrical symmetry. We also discuss the ambiguities in the form of the metric perturbation arising from gauge freedom and boundary conditions.
Using the geometric optics approximation, we then derive a set of coupled equations governing the propagation of electromagnetic and gravitational waves in the resulting curved spacetime. These equations are solved for two specific scenarios: a plane wave and a spherical wave. From the solutions, we compute the evolution of the wave amplitudes and the associated energy fluxes.
Our analysis reveals that two competing effects govern the amplitude evolution: magnification due to the focusing of waves by spacetime curvature, and attenuation due to wave conversion via the Gertsenshtein effect. In the plane wave case, these effects precisely cancel, resulting in no net change in amplitude. In contrast, for the spherical wave, the Gertsenshtein effect dominates over focusing, leading to an overall reduction in amplitude.
\end{abstract}

\maketitle

\section{Introduction}

The Gertsenshtein effect refers to the conversion of electromagnetic waves (EMWs) into gravitational waves (GWs) during their propagation through a background magnetic field (BGMF)~\cite{gertsenshtein1962wave}. This mechanism provides a possible means to constrain  background GWs ~\cite{Fujita2020Gravitational,Dolgov:2012be,Ito:2023nkq,Lella:2024dus,He:2023xoh}. The inverse process, known as the inverse Gertsenshtein effect, also occurs and  has been proposed as a possible method for detecting high-frequency gravitational waves~\cite{Boccaletti1970,Kolosnitsyn:2015zua,ejlliUpperLimitsAmplitude2019,Ito:2019wcb,Aggarwal:2020olq,Cruise:2012zz}.

In many theoretical treatments of Gertsenshtein effect, the BGMF is assumed weak, and the curvature it induces (BGMF-induced curvature) is often neglected ~\cite{gertsenshtein1962wave,Raffelt:1987im,DeLogi:1977qe,Palessandro2023Simple,Dolgov:2012be,Cembranos:2023ere,Ejlli:2020fpt}. The conversion probability — that is, the ratio of the energy flux of the incident EMWs to that of the converted GWs — is commonly calculated under this approximation. As stated in the literature, this probability is proportional to $B^2$, where $B$ is the magnetic field strength. On the other hand, the BGMF-induced curvature also arises at second order in $B$, and therefore contributes at the same order. Therefore, for a consistent treatment, the BGMF-induced curvature should be properly taken into account. Indeed, Ref.~\cite{Hwang2024GravitonPhoton} argues that in the vacuum Gertsenshtein effect, terms proportional to $B^2$ in the energy-momentum tensor, which were neglected in previous work, play a key role in the exponential creation of the converted wave. Additional terms proportional to $B^2$ coming from the BGMF-induced curvature were neglected in Ref.~\cite{Hwang2024GravitonPhoton}. Thus their conclusions may need to be reconsidered.

In summary, the treatment of the second order $B$ terms in the Gertsenshtein effect remains ambiguous and requires a rigorous formulation.
In this work, we derive the exact coupled equations for the Gertsenshtein effect up to second order in $B$ by considering the propagation of EMWs and GWs on the background metric sourced by a static and uniform magnetic field.

In our derivation, we adopt the geometric optics approximation. Geometric optics is a widely used approximation scheme based on the fundamental assumption that the wavelength is much smaller than all other characteristic length  scales, such as the curvature scale of the background spacetime or the curvature scale of the wavefront. This assumption generally holds for typical EMWs and GWs.

In Sec.~II, we provide the form of the background metric sourced by the BGMF and discuss the freedom in choosing it. In Sec.~III, we organize the equations describing the evolution of GWs and EMWs during propagation and establish the formalism of the graviton-photon system, exact up to order $B^2$. In Sec.~IV, we present solutions to the equations derived in Sec.~III and calculate the energy flux of EMWs in two specific cases.
Throughout this paper, we work in the natural unit $c=\hbar=1$.

\section{Determination of background metric}
\label{Sec:metric}
\begin{figure}[t]
  \begin{center}
    \includegraphics[clip,width=11.0cm]{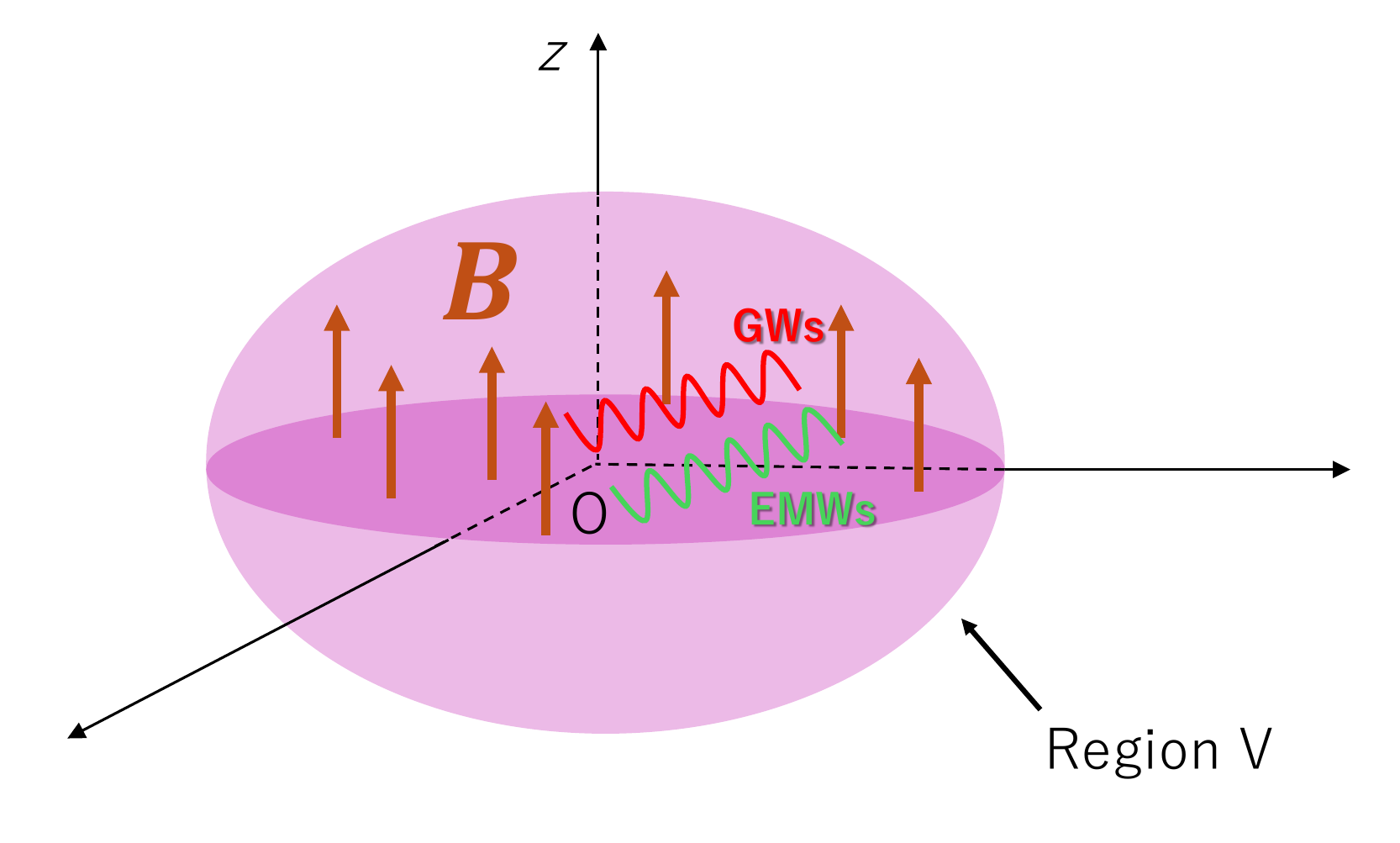}
    \caption{The region $V$ containing a uniform magnetic field along the $+z$ axis is depicted in purple.}
    \label{fig:fig1}
  \end{center}
\end{figure}

We consider a static and uniform magnetic field ${\bm B}$ in a finite volume region $V$. 
Without a loss of generality, we take the direction of the uniform magnetic field to be in the $+z$ direction,
\begin{align}
  \bm{B}=(0,0,B).
\end{align}
The energy-momentum tensor of this magnetic field (in a Cartesian coordinate system $x^\mu=(t,x,y,z$) is given by 
\begin{align}
\label{sec2:Tmunu-elemag}
  T_{\mu\nu}&=\frac{B^2}{2}\left( \begin{matrix}
1&0&0&0\\
0&1&0&0\\
0&0&1&0\\
0&0&0&-1
  \end{matrix}
   \right).
\end{align}
We assume that the size of the region $V$ is small enough that the spacetime sourced by the magnetic field 
${\bm B}$ deviates from the Minkowski spacetime only by a small amount. We also assume that the gravitational field is weak, i.e., that the size of the region $V$ is much larger than the Schwarzschild radius of the total energy of the magnetic field given by 
$\frac{B^2}{2}$ times the volume of the region $V$.
Then, we write the metric of such spacetime as 
\begin{align}
g^{(B)}_{\mu\nu}=\eta_{\mu\nu}+H_{\mu\nu},
\end{align}
where $H_{\mu\nu}$ is the part of the metric sourced by the uniform magnetic field $\bm{B}$.
We determine the metric $H_{\mu \nu}$ by solving the linearized Einstein equations.
The obtained metric $g^{(B)}_{\mu\nu}$, which is accurate up to ${\cal O}(B^2)$, 
serves as the background spacetime on which  EMWs and GWs propagate.
See Fig.~\ref{fig:fig1} for a schematic picture representing what we have described above.

The linearized Einstein equations in the Lorenz gauge $\partial^\mu {\bar H}_{\mu \nu}=0$
(${\bar H}_{\mu \nu}\equiv H_{\mu \nu}-\frac{1}{2}H\eta_{\mu \nu}$ and $H \equiv \eta^{\mu \nu}H_{\mu \nu}$)
read
\begin{align}
\triangle {\bar H}_{\mu \nu}=-16\pi G T_{\mu \nu}.
\end{align}
The static solution $H_{\mu \nu}$ that vanishes far away from the region $V$ is given by
\begin{align}
\label{sol-Lorenz-gauge}
{\bar H}_{\mu \nu} ({\bm x})=4G \int_V d^3x'~\frac{T_{\mu \nu} ({\bm x'})}{|{\bm x}-{\bm x'}|}.
\end{align}
Using the traceless nature of the electromagnetic energy-momentum tensor $T_{\mu \nu}$,  i.e., $\eta^{\mu \nu}T_{\mu \nu}=0$,
in the above solution, we have $\bar H=0$ and $H_{\mu \nu}={\bar H}_{\mu \nu}$.
Our main interest is the propagation of GWs and EMWs deep in the region $V$.
Thus, choosing the center of $V$ as the origin of our coordinate system and assuming  
reflection symmetry of the metric  for simplicity,
$H_{\mu \nu}$ near the origin may be written as
\begin{align}
H_{\mu \nu}({\bm x}) = a_{\mu \nu}+b_{\mu \nu i j}x^i x^j+\cdots,
\end{align}
where the coefficients $a_{\mu \nu}, b_{\mu \nu i j}$ may be obtained by expanding the solution
(\ref{sol-Lorenz-gauge}) around the origin.
Using the residual gauge $\triangle \xi_\mu=0$ to perform a further gauge transformation defined by
$\xi_\mu=\frac{1}{2}a_{\mu \nu}x^\nu$, the constant term $a_{\mu \nu}$ may be eliminated.
Thus, the metric around the origin can be expressed as
\begin{align}
\label{GB}g^{(B)}_{\mu\nu}=\eta_{\mu\nu}+b_{\mu \nu i j}x^i x^j.
\end{align}
As the coefficients $b_{\mu \nu i j}$ come from the original solution (\ref{sol-Lorenz-gauge}),
they depend on the shape of the boundary of $V$ as well as the behavior of $T_{\mu \nu}$ in
the vicinity of the boundary.
Note that $T_{\mu \nu}$, which is constant inside $V$ and zero outside of it, 
changes in a non-trivial manner near the boundary to satisfy the conservation law 
$\partial^\mu T_{\mu \nu}=0$.
As a result, even if only the diagonal components of $T_{\mu \nu}$ are non-vanishing, 
as given by Eq.~(\ref{sec2:Tmunu-elemag}) inside $V$, 
$H_{\mu \nu}$ generically exhibits non-vanishing off-diagonal components.
Thus, the expansion coefficients $b_{\mu \nu i j}$ are fully fixed only after
information at the boundary is given.
In this paper, we do not follow this approach.
Instead, we first define the metric around the origin as a solution of the Einstein equations
without making an explicit connection to the boundary, and then investigate the propagation of waves in this metric.

With this in mind, we go back to the original perturbed metric (\ref{GB}) (without imposing the Lorenz gauge)
and obtain the metric perturbation $H_{\mu \nu}$ as a solution of the linearized Einstein equations 
around the origin.
In the following, we assume that the system is static and has cylindrical symmetry around
the $z$-axis.
We employ a cylindrical coordinate system $(t,\rho,\varphi,z)$\footnote{In cylindrical coordinates \((t, \rho, \varphi, z)\), the Cartesian components \((t,x, y, z)\) are given by
$t=t,~x = \rho \cos\varphi,~y = \rho \sin\varphi,~z = z.$} and write the
perturbed metric as
\begin{align}
\label{sec2:cylindrical-metric}
ds^2=(-1+\Phi (\rho,z))dt^2+(1+\Psi (\rho,z))(d\rho^2+\rho^2 d\varphi^2 )+(1+\Omega (\rho,z) )dz^2.
\end{align}
Here, the metric functions $\Phi, \Psi, \Omega$ are ${\cal O}(B^2)$ and obey the linearized 
Einstein equations. Due to the cylindrical symmetry, the metric functions are independent of the azimuthal angle $\varphi$.
Substituting the above metric into the linearized Einstein equations, the non-trivial components become
\begin{align}
&2\Psi_{,zz}+\frac{1}{\rho} \Psi_{,\rho}+\frac{1}{\rho} \Omega_{,\rho}+\Psi_{,\rho \rho}+\Omega_{,\rho \rho}
=-8\pi G B^2 \\
&\Phi_{,zz}-\Psi_{,zz}+\frac{1}{\rho} \Phi_{,\rho}-\frac{1}{\rho} \Omega_{,\rho}=-8\pi GB^2 \\
&\Phi_{,\rho z}-\Psi_{,\rho z}=0, \\
&\Phi_{,zz}-\Psi_{,zz}+\Phi_{,\rho \rho}-\Omega_{,\rho \rho}=-8\pi GB^2, \\
&-\frac{1}{\rho}\Phi_{,\rho}+\frac{1}{\rho}\Psi_{,\rho}-\Phi_{,\rho \rho}+\Psi_{,\rho \rho}=-8\pi GB^2.
\end{align}
We impose that the metric reduces to the Minkowski metric $\eta_{\mu \nu}$ at the origin ($\rho=z=0$).
Then, the functions $\Phi, \Psi, \Omega$ near the origin may be expanded as
\begin{align}
\Phi =a_1\mathcal{R} \rho^2+a_2\mathcal{R} z^2,~~~\Psi=b_1\mathcal{R} \rho^2+b_2 Mz^2,~~~\Omega=c_1\mathcal{R} \rho^2+c_2\mathcal{R} z^2,
\end{align}
where for notational convenience we have defined a new quantity $\mathcal{R}$ with the dimensions of inverse length squared\footnote{$\mathcal{R} \simeq 2.5\times 10^{-37} \left({\frac{B}{1{\rm T}}}\right)^2 {\rm m}^{-2}$.} 
\begin{align}
\mathcal{R}=\frac{4\pi GB^2}{3} .
\end{align}
The expansion coefficients $a_1 \sim c_2$ are $\mathcal{O}(\mathcal{R}^0)$ constants.
Plugging these expansion coefficients into the above Einstein equations yields
the following three equations for the coefficients:
\begin{align}
\label{sec2:algebraic-eqs}
b_1+b_2+c_1=-\frac{3}{2},~~~a_1+a_2-b_2-c_1=-3,~~~
a_1-b_1=\frac{3}{2},
\end{align}
Thus, only three among the six coefficients are determined by the Einstein equations, leaving the remaining three as free parameters.
Notice that the following change of coordinates $(\rho,z)\to ({\bar \rho},{\bar z})$
\begin{align}
\rho = (1+\beta_1\mathcal{R} {\bar z}^2){\bar \rho},~~~z=(1-\beta_1 \mathcal{R}{\bar \rho}^2+\beta_2 \mathcal{R}{\bar z}^2 ){\bar z},  
\end{align}
where $\beta_1, \beta_2$ are arbitrary ${\cal O}(\mathcal{R}^0)$ constants, 
still keeps the form of the metric specified by Eq.~(\ref{sec2:cylindrical-metric}), and
only amounts to a shift in the coefficients,
\begin{align}
{\bar a_1}=a_1,~~~{\bar a_2}=a_2,~~~{\bar b_1}=b_1,~~~{\bar b_2}=b_2+2\beta_1,~~~{\bar c_1}=c_1-2\beta_1,~~~
{\bar c_2}=c_2+6\beta_2.
\end{align}
Thus, two among the three free parameters correspond to residual gauge degrees of freedom.
The remaining parameter is physical; different values of it correspond to different boundary conditions.
Solving the above algebraic equations (\ref{sec2:algebraic-eqs}) for $a_2, b_1, b_2$ gives
\begin{align}
a_2=-2a_1-3, ~~~~~b_1=a_1-\frac{3}{2},~~~~~b_2=-a_1-c_1.
\end{align}
To summarize, the perturbed metric sourced by the static and cylindrically symmetric uniform magnetic field 
can be written as
\begin{align}
\label{sec2:cylindrical-metric-2}
ds^2=\left( -1+\alpha \mathcal{R} \rho^2-(2\alpha+3)\mathcal{R}z^2 \right) dt^2+&\bigg[ 1+\left( \alpha-\frac{3}{2}\right)\mathcal{R} \rho^2 
-(\alpha+\beta_1) \mathcal{R}z^2 \bigg] (d\rho^2+\rho^2 d\varphi^2 ) \nonumber \\
&+(1+\beta_1 \mathcal{R}\rho^2+\beta_2 \mathcal{R}z^2 )dz^2,
\end{align}
where $\alpha$ (defined as $\alpha\equiv a_1$), $\beta_1, \beta_2$ are ${\cal O}(\mathcal{R}^0)$ arbitrary constants. 
The boundary conditions fix $\alpha$, and $\beta_1, \beta_2$ correspond to
gauge degrees of freedom. Explicit expressions of the gauge-invariant Riemann tensor (see Appendix~\ref{Appendix:R}) show that only $\alpha$ appears in it, providing another confirmation that
 $\alpha$ is a physically relevant quantity. In addition, $\alpha$ appears only as an overall factor in the Weyl tensor, suggesting that $\alpha$ parametrizes the strength of a tidal gravitational field generated by sources outside the region $r \ll \mathcal{R}^{-1/2}$ in which we expand the metric, due for example to the shape of the volume containing the magnetic field, or to variations in the strength of the magnetic field, or to other gravitational sources.

\section{Equations describing the photon-graviton system}
\subsection{Reduction of Einstein and Maxwell equations in geometric optics}\label{Amplitude}
Having established the background spacetime sourced by a static and uniform magnetic field, we now turn to deriving the equations governing the propagation of EMWs and GWs on this background. To this end, we decompose the total electromagnetic field $A^{\text{tot}}_\mu$ and the total metric $g^{\text{tot}}_{\mu\nu}$ into their background and perturbative components as
\begin{align}
g^{\text{tot}}_{\mu\nu} &= g^{(B)}_{\mu\nu} + h_{\mu\nu}, \label{decomposition-polarization-h} \\
A^{\text{tot}}_\mu &= A^{(B)}_\mu + A_\mu, \label{decomposition-polarization-A}
\end{align}
where the superscript $(B)$ denotes background quantities, and $h_{\mu\nu}$ and $A_\mu$ represent small perturbations corresponding to propagating EMWs and GWs, respectively.
Throughout this work, we treat these perturbations up to linear order
in the wave amplitude. 
Additionally, we assume that the wavelength of the waves, given by 
$\frac{2\pi}{\omega}$ (where $\omega$ denotes the angular frequency), is much smaller than 
the characteristic curvature scale $L$ of the background spacetime, 
or the curvature radius of the wavefront — whichever is smaller. 
Under this condition, the geometric optics approximation is applicable.

To derive the propagation equations under the geometric optics approximation, we express the perturbations $h_{\mu\nu}$ and $A_\mu$ as superpositions of different polarization modes:
\begin{align}
\label{reph}    
h_{\mu\nu} &= \sum_P \mathcal{H}_P \, \varepsilon_{\mu\nu}^P \, e^{i \phi}, \\
\label{repA}    
A_{\mu} &= \sum_P \mathcal{A}_P \, \varepsilon_{\mu}^P \, e^{i \phi},
\end{align}
where $P = 1, 2$ labels the polarization states. The quantities $\varepsilon_{\mu\nu}^P$ and $\varepsilon_\mu^P$ denote the polarization tensor and vector corresponding to the gravitational and electromagnetic perturbations, respectively. These polarization bases are normalized as
\begin{align}
\varepsilon_{\mu\nu}^P \varepsilon^{\mu\nu}_{P'} = 2 \delta^P_{~P'}, \qquad
\varepsilon_\mu^P \varepsilon^\mu_{P'} = \delta^P_{~P'}.
\end{align}
The phase $\phi$ varies on the scale of the wavelength, i.e., $|\partial_\mu \phi|/|\phi|=\mathcal{O}(\omega)$, while the amplitudes $\mathcal{H}_P$ and $\mathcal{A}_P$ vary on the much larger scale $L$.

Making use of the invariance of the wave equations under the following gauge transformations,
\begin{equation}
h_{\mu \nu} \to h_{\mu \nu} - \xi_{\mu ; \nu} - \xi_{\nu ; \mu}, \qquad
A_{\mu} \to A_{\mu} - \chi_{,\mu},
\end{equation}
where $\xi_\mu$ and $\chi$ are arbitrary functions
and ``$;$'' denotes the covariant derivative with respect to the background metric $g^{(B)}_{\mu\nu}$, 
we impose the following gauge conditions on the perturbations $h_{\mu \nu}$ and $A_\mu$:
\begin{align}
\label{gaugeh}
h_{\mu\nu}{}^{;\mu} &= 0, \\
h_\mu{}^\mu&=0,\\
\label{gaugeA}
A_\mu{}^{;\mu} &= 0.
\end{align}
Substituting the mode expansions in Eqs.~\eqref{reph} and \eqref{repA}, 
and working to leading order in the small expansion parameter ${(L \omega)}^{-1}$, these gauge conditions reduce to
\begin{equation}
P^\mu \varepsilon_{\mu \nu}^P = 0, \qquad \varepsilon^P_\mu{}^\mu = 0 , \qquad P^\mu \varepsilon_\mu^P = 0,
\end{equation}
where $P^\mu \equiv \partial^\mu \phi$ is the wave vector normal to hypersurfaces of constant phase $\phi$.
In geometric optics, waves are described by a congruence of rays whose 
tangent vectors are given by $P^\mu$.

With the above decomposition and gauge conditions in place, 
we are now ready to derive the propagation equations for the mode amplitudes 
$\mathcal{H}_P$ and $\mathcal{A}_P$ from the linearized Einstein and Maxwell equations.
The Einstein equations at first order in perturbation under the gauge conditions are given by
\begin{align}
\label{linearized-Ein-hA}
-h_{\mu\nu;\alpha}{}^{\alpha}=16\pi G \delta T_{\mu\nu},
\end{align}
where $\mathcal{O}({\omega}^0)$ terms have been dropped and $\delta T_{\mu \nu}$ is the energy-momentum tensor of the electromagnetic field at
first order in $A_\mu$, given by
\begin{align}
\delta T_{\mu \nu}=F_{~~~\mu}^{(B)\alpha} \delta F_{\nu \alpha}+F_{~~~\nu}^{(B)\alpha} \delta F_{\mu \alpha}
-\frac{1}{2} g^{(B)}_{\mu \nu} F_{(B)}^{\alpha \beta} \delta F_{\alpha \beta},
\hspace{15mm}
\delta F_{\alpha \beta} \equiv \partial_\alpha A_\beta-\partial_\beta A_\alpha.
\end{align}
Substituting the decomposition (\ref{decomposition-polarization-h}) and (\ref{decomposition-polarization-A}),
the left-hand side of the Einstein equations (\ref{linearized-Ein-hA}) up to ${\cal O}(\omega)$ becomes
\begin{align}
-h_{\mu\nu;\alpha}{}^{\alpha}
=\sum_{P'} \bigg[ P_\alpha P^\alpha \mathcal{H}_{P'} \varepsilon^{P'}_{\mu \nu}
-i (2 P^\alpha \varepsilon^{P'}_{\mu \nu ; \alpha}+2 P^\alpha \mathcal{H}_{P';\alpha} \varepsilon^{P'}_{\mu \nu}
+P^\alpha{}_{;\alpha} \mathcal{H}_{P'} \varepsilon^{P'}_{\mu \nu})
\bigg] e^{i\phi}+\mathcal{O}(\omega^0) .
\end{align}
Similarly, the right-hand side becomes
\begin{align}
16\pi G \delta T_{\mu\nu}=16\pi G i \sum_{P'} \bigg[ F_{~~~\mu}^{(B)\alpha}
(\varepsilon^{P'}_\alpha P_\nu -\varepsilon^{P'}_\nu P_\alpha)+
F_{~~~\nu}^{(B)\alpha} (\varepsilon^{P'}_\alpha P_\mu -\varepsilon^{P'}_\mu P_\alpha)
-\frac{1}{2} g^{(B)}_{\mu \nu} F_{(B)}^{\alpha \beta}
(\varepsilon^{P'}_\beta P_\alpha-\varepsilon^{P'}_\alpha P_\beta)
\bigg]e^{i\phi}+\mathcal{O}(\omega^0).
\end{align}
Plugging these expressions to Eq.~(\ref{linearized-Ein-hA})
and contracting the resultant equations with the polarization tensor 
$\varepsilon_{P}^{\mu \nu}$,
we obtain
\begin{align}
2 \bigg[ P_\alpha P^\alpha \mathcal{H}_P-i
( 2P^\alpha \mathcal{H}_{P;\alpha}+P^\alpha{}_{;\alpha}\mathcal{H}_P ) \bigg] =-32\pi Gi 
\sum_{P'} \mathcal{A}_{P'} F^{(B)}_{\mu \nu}P^\nu \varepsilon^{P'}_\alpha
\varepsilon_P^{\mu \alpha}.
\end{align}
Requiring this equation to hold at each order in $\omega$ yields the
following two equations;
\begin{align}
&P_\alpha P^\alpha =0, \label{null-eq} \\
&P^\alpha \mathcal{H}_{P;\alpha}+\frac{P^\alpha{}_{;\alpha}}{2} \mathcal{H}_P=8\pi G 
\sum_{P'} \mathcal{A}_{P'} F^{(B)}_{\mu \nu}P^\nu \varepsilon^{P'}_\alpha
\varepsilon_P^{\mu \alpha}. \label{HP-eq}
\end{align}
Using $P_\mu=\partial_\mu \phi$, the first equation (\ref{null-eq}) gives
\begin{align}
\label{GWs-geodesic-eq}
 P^\alpha P_{\mu ; \alpha}=0.
\end{align}
Thus, the trajectories of GWs, which have tangent vector $P^\mu$, obey the (null) geodesic equation.
The second equation (\ref{HP-eq}) describes how the GW amplitude changes as the waves propagate
along the geodesic specified by $P^\mu$.
The second term on the left-hand side of this equation, proportional to the term of $P^\mu{}_{;\mu}$, represents the magnification/de-
magnification effects arising from the convergence and divergence of geodesics.
The factor $\frac{1}{2}$ reflects the fact that the square of the wave amplitude ($\propto$ flux) 
is inversely proportional to the cross-sectional area  of the null congruence.
The right-hand side, which appears only when the background magnetic field is present,
shows the Gertsenshtein effect (sometimes called photon-graviton conversion), 
in which gravitational waves are generated 
out of the EMWs through the static magnetic field.
This term depends on the background magnetic field through the combination $F_{\mu \nu}^{(B)}P^\nu$,
implying that only the component of ${\bm B}$ perpendicular to the propagation direction of EMWs
enters the Gertsenshtein effect.
As it is obvious from the derivation of the equation, the frequency of the GWs
converted from the EMWs is the same as the frequency of the EMWs.

Plugging Eqs.~(\ref{null-eq}) and (\ref{HP-eq}) back into the linearized Einstein equations,
we obtain the evolution equation for the polarization tensor
\begin{align}
P^\alpha \varepsilon^{P}_{\mu \nu ; \alpha}=0. \label{polarization-tensor-eq}
\end{align}
Thus, in the geometric optics approximatoin, the polarization tensor $\varepsilon^P_{\mu \nu}$ is parallel-transported along
the null geodesics.

To summarize, in geometric optics, the linearized Einstein equations reduce to
Eqs.~(\ref{HP-eq}), (\ref{GWs-geodesic-eq}), and (\ref{polarization-tensor-eq}),
which determine $\mathcal{H}_{P}$, $P^\mu$, and $\varepsilon^{P}_{\mu \nu}$, respectively.

Similarly, the Maxwell equations at first order in perturbation
under the gauge conditions above are given by
\begin{align}
A_{\mu; \alpha}^{~~~;\alpha}-F^{\alpha}_{(B) \beta} \delta \Gamma^\beta_{~\alpha \mu}=0,
\end{align}
where $\delta \Gamma^\beta_{~\alpha \mu}$ is the Christoffel symbol to first order in the GW amplitude,
and $\mathcal{O}(\omega^0)$ terms have been dropped.
Plugging the decomposition (\ref{decomposition-polarization-h}) and (\ref{decomposition-polarization-A}),
we obtain
\begin{align}
\label{linearized-Maxwell-eqs}
\sum_{P'} \bigg[ -P_\alpha P^\alpha \mathcal{A}_{P'} \varepsilon^{P'}_\alpha 
+i (2P^\alpha \mathcal{A}_{P';\alpha}+P^\alpha{}_{;\alpha} \mathcal{A}_{P'}) \varepsilon^{P'}_\mu
+2i P^\alpha\varepsilon^{P'}_{\mu ; \alpha} \mathcal{A}_{P'} \bigg] e^{i\phi}+
i F_{(B)}^{\alpha \beta} \sum_{P'} \mathcal{H}_{P'} \varepsilon^{P'}_{\alpha \mu}
P_\beta e^{i\phi}=0.
\end{align}
Contracting this equation with the polarization vector $\varepsilon_{P}^\nu$,
we obtain
\begin{align}
 -P_\alpha P^\alpha \mathcal{A}_{P}
+i (2P^\alpha \mathcal{A}_{P;\alpha }+P^\alpha{}_{;\alpha} \mathcal{A}_{P}) 
+i F_{(B)}^{\alpha \beta} \sum_{P'} \mathcal{H}_{P'} \varepsilon^{P'}_{\alpha \mu} \varepsilon_P^\mu
P_\beta=0.
\end{align}
Requiring this equation to hold at each order in $\omega$
 yields the
following two equations;
\begin{align}
&P_\alpha P^\alpha =0, \label{null-eq-EMW} \\
&P^\alpha \mathcal{A}_{P;\alpha}+\frac{P^\alpha{}_{;\alpha}}{2} \mathcal{A}_P=
-\frac{1}{2} \sum_{P'} \mathcal{H}_{P'} F_{(B)}^{\alpha \beta} \varepsilon^{P'}_{\alpha \mu} \varepsilon_P^\mu
P_\beta. \label{AP-eq}
\end{align}
Again, the first equation yields the geodesic equation (\ref{GWs-geodesic-eq}).
Analogous to the case of GWs, the second equation describes the evolution of the EMW amplitude along the geodesic, with the second term on the left-hand side representing the same effect as in the GW case.
The right-hand side corresponds to the inverse Gertsenshtein effect, conversion
of GWs into EMWs through the background magnetic field.
This effect is also sensitive only to the component of ${\bm B}$ perpendicular to the propagation direction of GWs.
Plugging the above two equations back into Eqs.~(\ref{linearized-Maxwell-eqs}) gives the
evolution equation for the polarization vector
\begin{align}
\label{polarization-vector-eq}
P^\mu \varepsilon^{P}_{\nu ; \mu}=0.
\end{align}
Thus, the polarization vector $\varepsilon^P_{\mu}$ is parallel-transported along
the null geodesics.
To summarize, in geometric optics, the Maxwell equations reduce to
Eqs.~(\ref{AP-eq}), (\ref{GWs-geodesic-eq}), and (\ref{polarization-vector-eq}),
which determine $\mathcal{A}_{P}$, $P^\mu$, and $\varepsilon^{P}_\mu$, respectively.

\subsection{Construction of the polarization vector and tensor}
\label{def:bases}
In order to solve the equations for the wave amplitudes (\ref{HP-eq}) and (\ref{AP-eq}),
we need to specify a basis for the polarization states.
In this work, we consider an infinitesimal bundle of null geodesics representing a ``beam" of GWs and EMWs having the same frequency and emanating from the origin at an angle $\theta$ measured from the $+z$--axis. 

\begin{figure}[t]
  \begin{center}
    \includegraphics[clip,width=11.0cm]{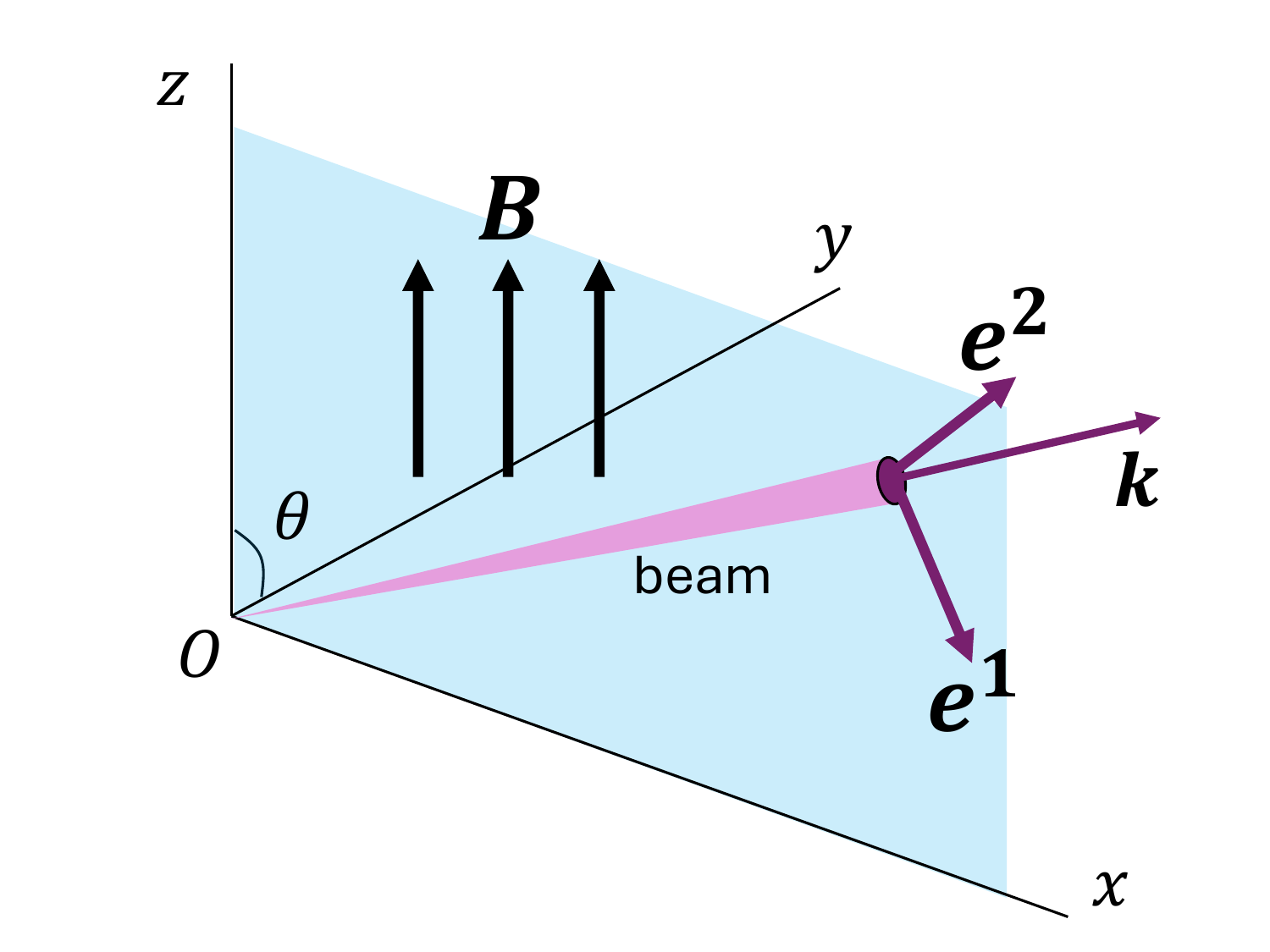}
    \caption{A narrow beam emanating from the origin moves in the $x$--$z$ plane with an initial angle
    $\theta$ from the $+z$--axis. The three-dimensional vector ${\bm k}$ which is the spatial 
    component of $P^\mu$ represents the direction of the beam.}
    \label{fig:null-geodesic}
  \end{center}
\end{figure}

We take the $x$--$z$ plane so that the central geodesic in the beam is confined to that plane, 
as illustrated schematically in Fig.~\ref{fig:null-geodesic}.
This setup can be adopted without loss of generality, provided that the background magnetic field and spacetime possess cylindrical symmetry about the $z$--axis.
Then, the vector $P^\mu=\partial^\mu \phi$ at the origin becomes $P^\mu=\omega_0 (1,\sin \theta, 0, \cos \theta)$ in
the Cartesian coordinates, where $\omega_0$ is the angular frequency
of the waves as measured by an observer at the origin at rest in the $(t,x,y,z)$ coordinate system. Under this convention, \( P^\mu \) represents the four-momentum of a single photon or graviton in the particle picture, where EMWs and GWs are regarded as ensembles of photons and gravitons, respectively. By using $P^\mu$, the angular frequency observed by an observer with four-velocity \( u^\mu \) is given by
\begin{equation}
    \omega_{\text{obs}} = -P^\mu u_\mu.
\end{equation}

For later convenience, we define a dimensionless null tangent vector \( k^\mu \) by
\begin{equation}
    k^\mu \equiv \frac{P^\mu}{\omega_0}.
\end{equation}
This \( k^\mu \) still satisfies the geodesic equation, and its solutions are the same as those for \( P^\mu \). Therefore, in the following discussion, we describe the trajectory of a beam using \( k^\mu \) instead of \( P^\mu \). Moreover, we introduce the affine parameter \( \lambda \) to parametrize the central geodesic \( x^\mu(\lambda) \), and set $ \lambda = 0 $ at the origin.  
In this convention, \( \lambda \) has dimensions of time in natural units.

At the origin ($\lambda=0$), we decompose the polarization vectors  into a standard linear-polarization basis consisting of two space-like unit vectors normal to each other and to the spatial direction of the central geodesic.
The components of these vectors
in the Cartesian coordinates $(t,x,y,z)$ defined earlier are
\begin{align}
\label{Initial e}
e^1_{\mu} (\lambda=0) = (0, \cos\theta, 0, -\sin\theta),\qquad
e^2_{\mu} (\lambda=0) = (0, 0, 1, 0).
\end{align}
The vectors $e^1_{\mu} (\lambda=0)$ and $e^2_{\mu} (\lambda=0)$ are then parallel transported along the geodesics 
to define a new basis $e^1 (\lambda), e^2(\lambda)$ at $x=x^\mu(\lambda)$. Its explicit expression is given in Appendix \ref{appendix:tetrad}.
Because $e^1_\mu k^\mu=0$ and $e^2_\mu k^\mu=0$ are satisfied along
the geodesic, $e^1_\mu (\lambda)$ and $ e^2_\mu (\lambda)$ are still perpendicular to the geodesic.
For later convenience, we define the plane $\Sigma(\lambda)$ as a plane spanned by $e^1_\mu (\lambda)$ and $ e^2_\mu (\lambda)$.
Thus, $e^1_\mu (\lambda), e^2_\mu (\lambda)$ form an orthonormal basis on the plane $\Sigma(\lambda)$. We give the cross-section area of the beam at $\lambda$ as the cross-section of the beam on $\Sigma(\lambda)$.
Also, because of the transverse nature of the GWs and the EMWs,
the polarization tensors and vectors are defined on this plane. 

Having established the basis that spans the cross-sectional area of the beam, the next task is to derive expressions of the polarization tensors and vectors satisfying Eqs.~(\ref{polarization-tensor-eq}) and (\ref{polarization-vector-eq}) in terms of the specified basis.
Actually, they are given by
\begin{align}
\varepsilon^1_{\mu\nu} (\lambda)& =e^{1}_\mu (\lambda)\otimes e^{2}_\nu (\lambda)+ e^{2}_\mu (\lambda)\otimes e^{1}_\nu(\lambda) , \\
\varepsilon^2_{\mu\nu}(\lambda)&=e^{2}_\mu (\lambda) \otimes e^{2}_\nu(\lambda) - e^{1}_{\mu}(\lambda) \otimes e^{1}_{\nu}(\lambda) , \\
\varepsilon^1_{\mu}(\lambda)&= e^{1}_\mu (\lambda), \\
\varepsilon^2_{\mu}(\lambda)&=e^{2}_\mu(\lambda).
\end{align}
The polarization states of gravitational waves with $P=1,2$ correspond to the standard cross and plus modes, respectively. 
Note that for convenience we define $\varepsilon^{2}_{\mu\nu}$ with a sign opposite to the standard one.
In the following subsection, we derive the equations governing the evolution of GW and EMW 
amplitudes during the propagation by using these polarization tensors and vectors.

\subsection{Equations for the wave amplitudes}
\label{subsec:wave-amplitude}
The right-hand sides of both \eqref{HP-eq} and \eqref{AP-eq} contain the term
$F_{(B)}^{\alpha \beta}  \varepsilon^{P'}_{\alpha \mu} \varepsilon_P^\mu P_\beta$. Using the expressions of the polarization tensors and vectors defined in the previous 
subsection, this term can be written as
\begin{align}
\label{eq:Feek}
    F_{(B)}^{\alpha \beta}  \varepsilon^{P'}_{\alpha \nu} \varepsilon_P^\nu
P_\beta=-B\omega_0\sin \theta \delta^{P'}_{P}.
\end{align}
The presence of the Kronecker delta $\delta^{P'}_{P}$ indicates that the cross mode (resp., plus mode)
of GWs couples to EMWs linearly polarized along the $e^1_{\mu}$ (resp., $e^2_{\mu}$) direction.
Equivalently, in terms of circular polarization,
the left-handed (right-handed) polarization mode of GWs is coupled to the left-handed (right-handed) polarization mode of EMWs, and modes with different circular polarization are decoupled.
Plugging Eq.~(\ref{eq:Feek}) into \eqref{HP-eq} and \eqref{AP-eq}, the equations 
for $\mathcal{H}^{P}$ and $\mathcal{A}^{P}$ become
\begin{align}
\label{H-evo}\frac{d \mathcal{H}^{P}}{ d\lambda} &=-\frac{\Theta}{2} \mathcal{H}^{P} + 8\pi G\, \mathcal{A}^{P} B \sin\theta, \\
\label{A-evo}\frac{d\mathcal{A}^{P}}{d\lambda}  &= -\frac{\Theta}{2} \mathcal{A}^{P}  -\frac{1}{2} \mathcal{H}^{P} B \sin\theta,
\end{align}
where $\Theta\equiv k^\mu{}_{;\mu}$ is the expansion of the null congruence.
These equations show that the forms of the evolution equations for $\mathcal{H}^{P}$ and $\mathcal{A}^{P}$ do not depend on the polarization state.
Because of this property, in the following analysis, we omit the index ``P''.

It is useful to note at this stage that Eqs.~(\ref{H-evo}) and (\ref{A-evo}) imply the conservation law $j^\mu{}_{;\mu}=0$, where the current is defined as $j^\mu\equiv\qty(\mathcal{A}^2+\frac{1}{\gamma^2}\mathcal{H}^2)P^\mu$. 
Here, $\gamma \equiv \sqrt{16\pi G}$ appears with $\mathcal{H}$ 
because $\gamma^{-1} \mathcal{H}$  is the canonically normalized field.
Adopting the view that EMWs and GWs are ensembles of photons and gravitons, 
$\omega_{\text{obs}} \mathcal{A}^2$ and $\omega_{\text{obs}}\frac{\mathcal{H}^2}{\gamma^2}$ represent the number densities
of photons and gravitons, respectively \footnote{Since photons and gravitons following a given null geodesic are moving in the same direction, $\omega_{\rm obs} \mathcal{A}^2$ and $\omega_{\text{obs}}\frac{\mathcal{H}^2}{\gamma^2}$ are also the number fluxes of photons and gravitons, respectively.}\label{footnote-numberflux}.
The conservation law thus expresses the conservation of the total number of photons
and gravitons.
This is a natural consequence given that the background magnetic field only acts in
changing photons to gravitons or vice versa.

Because the expansion $\Theta$ appears in Eqs.~\eqref{H-evo} and \eqref{A-evo},
the evolution of $\mathcal{H}$ and $\mathcal{A}$ as the solution of Eqs.~\eqref{H-evo} and \eqref{A-evo} can be determined only after $\Theta$ is given.
Since $\Theta$ depends on the properties of the congruence, it cannot be derived solely from the geodesic equation \eqref{GWs-geodesic-eq}. 
In the following subsection, we present the evolution equation of $\Theta$ and other physical quantities characterizing the congruence such as the shear and the rotation. 

\subsection{Equations for the expansion, shear, and rotation} 
The evolution of the congruence is characterized by the tensor field 
$B_{\mu\nu} = k_{\mu ;\nu}$. 
If the congruence is composed of null geodesics, the physically meaningful part of $k_{\mu ;\nu}$ is its
projection onto the two-dimensional planes $\Sigma(\lambda)$.
Denoting the projection operator by $Q_{\mu\nu}$\footnote{$Q_{\mu\nu}$  can be written explicitly as $Q_{\mu\nu}=g_{\mu\nu}+k_{\mu}n_{\nu}+n_{\mu}k_{\nu}$, where $n^\mu$ is the axial vector which satisfies
$n^\mu k_\mu = -1$, $n^\mu n_\mu = 0$, and is orthogonal to the plane $\Sigma(\lambda)$. Alternatively, using the basis defined in Sec.~\ref{def:bases}, $Q_{\mu\nu}$  also can be written as $Q_{\mu\nu}=e_{\mu}^1\otimes e_{\nu}^1\otimes+  e _{\mu}^2\otimes  e _{\nu}^2$.},
the projection of $k_{\mu ;\nu}$ onto $\Sigma$ is given by $\hat{B}_{\mu\nu}\equiv P_{\alpha;\beta}Q^{\alpha}{}_{\mu}Q^{\beta}{}_{\nu}$.
We use the symbol $\hat{}$ to indicate that the corresponding quantity is projected onto $\Sigma(\lambda)$ using $Q_{\mu\nu}$.
Then, $\hat{B}_{\mu\nu}$ is decomposed as~\cite{carroll2004spacetime}
\begin{align}
\hat{B}_{\mu\nu}=\frac{1}{2}\Theta Q_{\mu\nu}+\hat{\sigma}_{\mu\nu}+\hat{\omega}_{\mu\nu}.
\end{align}
Here, the congruence expansion $\Theta$, shear $\hat{\sigma}_{\mu\nu}$, and rotation $\hat{\omega}_{\mu\nu}$, are defined by
\begin{align}
\label{Expansion:def}\Theta&=k_{\mu}{}^{;\mu}=\hat{B}_{\mu\nu}Q^{\mu\nu},\\
\label{Shear:def}\hat{\sigma}_{\mu\nu}&=\hat{B}_{(\mu\nu)}-\frac{1}{2}\Theta Q_{\mu\nu},\\
\label{rotation:def}\hat{\omega}_{\mu\nu}&=\hat{B}_{[\mu\nu]}.
\end{align}
These quantities characterize the deformation of the cross-sectional area of the congruence. 
The expansion $\Theta$ describes the isotropic expansion or contraction of the cross-sectional area, whereas the shear ${\hat \sigma}_{\mu\nu}$ describes anisotropic shape deformations that preserve the cross-sectional area.

The evolution equations of the expansion and the shear are given by~\cite{carroll2004spacetime}
\begin{align}
   \label{Expansion} \frac{d\Theta}{d\lambda}&=-\frac{1}{2}\Theta^2-\hat{\sigma}^{\mu\nu}\hat{\sigma}_{\mu\nu}-R_{\mu\nu}k^\mu k^\nu,\\
  \label{Shear}  \frac{D\hat{\sigma}_{\mu\nu}}{d\lambda}&=-\Theta\hat{\sigma}_{\mu\nu}-Q^\alpha{}_{\mu}Q^{\beta}{}_{\nu}C_{\alpha\rho\beta\sigma}k^\rho k^\sigma,
\end{align}
where the Riemann tensor \( R^\mu{}_{\nu\rho\sigma} \) is defined by
\( 2 V_\mu{}_{;[\rho\sigma]} = R^\nu{}_{\mu\rho\sigma} V_\nu \) for any vector \( V^\mu \), the Ricci tensor \( R_{\mu\nu} \) is defined as \( R_{\mu\nu} = R^\alpha{}_{\mu\alpha\nu} \), and the Weyl tensor \( C_{\rho\sigma\mu\nu} \) is defined by $C_{\rho\sigma\mu\nu} = R_{\rho\sigma\mu\nu} - \frac{1}{2} \left( g_{\rho[\mu} R_{\nu]\sigma} - g_{\sigma[\mu} R_{\nu]\rho} \right) + \frac{1}{6} R\, g_{\rho[\mu} g_{\nu]\sigma}.$ And $\frac{D}{d\lambda}$  is the covariant derivative along  $P^\mu$.

In our study, $R_{\mu \nu}$ and $C_{\mu\nu\lambda\sigma}$ appearing in the above equations
should be evaluated using the background metric $g^{(B)}_{\mu \nu}$.
Once the initial conditions for $\Theta$ and $\hat{\sigma}_{\mu\nu}$ are given,
which is equivalent to specifying the initial configuration of the beams,
$\Theta$ and $\hat{\sigma}_{\mu\nu}$ at arbitrary $\lambda$ can be obtained by solving
the above equations.
Then, $\Theta$ obtained in this way is used to finally determine $\mathcal{H}$ and $\mathcal{A}$
as the solution of Eqs.~\eqref{H-evo} and \eqref{A-evo}.
In appendix~\ref{Appendix:visualCS}, we solve the geodesic deviation equations directly
to evaluate the deformation of the cross-section of the null congruence, from which
the evolution of $\Theta$ is determined.

\section{Evolution of the energy flux of EMWs and GWs during propagation}\label{Sec.solution}
In this section, based on the basic equations derived in the previous section, we investigate the evolution of the energy fluxes of EMWs and GWs, denoted as \( E_{\text{EMWs}} \) and \( E_{\text{GWs}} \), respectively. 
As mentioned in footnote in \ref{footnote-numberflux},
if EMWs and GWs are viewed as ensembles of photons and gravitons,
$\mathcal{A}^2 \omega_{\text{obs}}$ and $ \frac{\mathcal{H}^2}{\gamma^2} \omega_{\text{obs}}$ may be understood as the number flux of photons and gravitons, respectively.
Thus, the energy fluxes of photons and gravitons, or equivalently the energy fluxes of EMWs and GWs, as measured by an observer with four-velocity \( u^\mu \), are given by
\begin{align}
    E_{\text{EMWs}} &= \mathcal{A}^2 P^\mu P^\nu u_\mu u_\nu = \mathcal{A}^2 \omega_{\text{obs}}^2, \label{eq:EMW_energy} \\
    E_{\text{GWs}}&= \frac{\mathcal{H}^2}{\gamma^2} P^\mu P^\nu u_\mu u_\nu =  \frac{\mathcal{H}^2}{\gamma^2} \omega_{\text{obs}}^2. \label{eq:GW_energy}
\end{align}

As these expressions indicate, the energy fluxes depend on the observer, since the observed angular frequency $\omega_{\text{obs}}$ varies with the observer's motion. To make this dependence more explicit, let us consider two representative observers who measure the energy flux at the spacetime point $x^\mu(\lambda)$.
In the first case, the observer's four-velocity $u^\mu$ at the measurement point is defined by parallel transporting the vector $(1,0,0,0)$,
which is given at the origin, along the null geodesic $x^\mu(\lambda)$. 
For this observer, the measured angular frequency is
\begin{equation}
    \omega_{\text{obs}}=\omega_0,
\end{equation}
which means the observed frequency remains unchanged from the initial frequency $\omega_0$.
In the second example, the observer is associated with the timelike Killing vector field that characterizes the static nature of the spacetime. For this observer, the measured angular frequency is given by
\begin{equation}
    \omega_{\text{obs}}= \omega_0 \sqrt{-g^{00}_{(B)}} 
    = \omega_0 
    \qty(1 + \frac{\mathcal{R} \omega_0^2\lambda^2}{2 } \qty[-3(1 + \alpha)\sin^2\theta + 2\alpha-3]).
\end{equation}
This observer detects a gravitational redshift or blueshift. 
These examples explicitly shows how the frequency—and thus the energy flux—depends on the observer's frame, particularly through the spacetime geometry encoded in $g^{00}_{(B)}$.

As discussed in the previous section, the evolution of the wave amplitudes depends on the initial values of $\Theta$ and $\hat{\sigma}_{\mu \nu}$, namely, 
on the configuration of the beams. 
In this study, we consider two simple and representative beam configurations: one corresponding to a plane wave, and the other to a spherical wave.
We first derive the general solution for \( \mathcal{A}^2 \) and \( \mathcal{H}^2 \), which explicitly contain \( \Theta \), from Eqs.~\eqref{H-evo} and \eqref{A-evo}. 
Then, for each case, we compute \( \Theta \) using Eqs.~\eqref{Expansion} and \eqref{Shear}. 
Finally, we substitute it into the general solutions, obtain the resulting 
evolution of \( \mathcal{A}^2 \) and \( \mathcal{H}^2 \) as functions of 
$\lambda$, and discuss the implications of the results.

\subsection{General solution of Eqs.~\eqref{H-evo} and \eqref{A-evo}}

Solving Eqs.~\eqref{H-evo} and \eqref{A-evo}, the general solution in terms of 
\( \mathcal{A}^2\) and \( \mathcal{H}^2 \) can be written as follows:
\begin{align}
\label{AMPH}
\mathcal{H}^2(\lambda) &= \left[
    \mathcal{H}_0 \exp\left( -\frac{1}{2} \int_0^\lambda \Theta \, d\lambda' \right)
    \cos\left( \frac{1}{2} \gamma B \lambda \sin\theta \right)
    + \gamma\mathcal{A}_0 \exp\left( -\frac{1}{2} \int_0^\lambda \Theta \, d\lambda' \right)
    \sin\left( \frac{1}{2} \gamma B \lambda \sin\theta \right)
\right]^2, \\
\label{AMPA}
\mathcal{A}^2(\lambda) &= \left[
    -\frac{\mathcal{H}_0}{\gamma} \exp\left( -\frac{1}{2} \int_0^\lambda \Theta \, d\lambda' \right)
    \sin\left( \frac{1}{2} \gamma B \lambda \sin\theta \right)
    + \mathcal{A}_0 \exp\left( -\frac{1}{2} \int_0^\lambda \Theta \, d\lambda' \right)
    \cos\left( \frac{1}{2} \gamma B \lambda \sin\theta \right)
\right]^2,
\end{align}
where
$\mathcal{H}_0$ and $\mathcal{A}_0$ represent the amplitudes of GWs and EMWs, respectively, at $\lambda = 0$. 
Note that since the previous equations \eqref{H-evo} and \eqref{A-evo} are valid up to order $\mathcal{O}(B^2)$, the solution above is also valid up to $\mathcal{O}(B^2)$. 

In the following subsections, we restrict our analysis to cases
where only EMWs are present at $\lambda = 0$, i.e., $\mathcal{H}_0 = 0$
and GWs completely originate from the Gertsenshtein effect. 
In this case, the solutions of Eqs~\eqref{AMPH} and \eqref{AMPA} can be simplified as follows:
\begin{align}
\label{AMPH10nougawA_expanded}
\mathcal{H}^2(\lambda) &= 
\gamma^2 \mathcal{A}_0^2 
\exp\left( -\int_0^\lambda \Theta(\lambda') \, d\lambda' \right)
\sin^2\left( \frac{1}{2} \gamma B\lambda \sin\theta \right), \\
\label{AMPA_expanded}
\mathcal{A}^2(\lambda) &= 
\mathcal{A}_0^2 
\exp\left( -\int_0^\lambda \Theta(\lambda') \, d\lambda' \right)
\cos^2\left( \frac{1}{2} \gamma B \lambda \sin\theta \right).
\end{align}

In the evolution equation for \( \mathcal{H}^2 \), since the leading term is \( \mathcal{O}(B^2) \), the \( \mathcal{O}(B^2) \) correction induced from BGMF-induced curvature does not affect the leading-order behavior. In contrast, the leading term of \( \mathcal{A}^2 \) is \( \mathcal{O}(B^0) \), so the \( \mathcal{O}(B^2) \) correction from \( \Theta \) plays a significant role in the evolution of \( \mathcal{A}^2 \).

On the other hand, since the evolution of the EMW energy flux during propagation is influenced by the BGMF-curvature, its effect should be properly included.

In closing this subsection, it is helpful to compare our results with 
previous work on the photon-graviton conversion probability.
In quantum mechanics, the waves are interpreted as collections of particles and the number flux is given by the energy flux divided by the wave frequency: $N_\gamma (\lambda)\equiv E_{\text{EMWs}}/\omega_{\rm{obs}},~ N_g(\lambda) \equiv E_{\text{GWs}}/\omega_{\rm{obs}}$. 
The conversion probability of photons into gravitons is then identified with
\begin{equation}
P_{\gamma\to g}\equiv\frac{N_g(\lambda)}{N_\gamma(\lambda)+N_g(\lambda)}=\frac{E_{\text{GWs}}}{E_{\text{EMWs}}+E_{\text{GWs}}}
\end{equation}
for $N_g(\lambda=0)=0$. 
Using the solutions (\ref{AMPH10nougawA_expanded}) and (\ref{AMPA_expanded}) for Eqs.~(\ref{eq:EMW_energy}) and (\ref{eq:GW_energy}), we have
\begin{equation}
    P_{\gamma\to g}=\sin^2\qty(\frac{1}{2}\gamma B\lambda\sin \theta)=4\pi GB^2\lambda^2\sin^2\theta +{\cal O}(B^4).
    \label{conversionprobability}
\end{equation}
Notice that $\omega_{\rm{obs}}$ does not appear in the final expression, meaning that the conversion probability is 
independent of the observer's motion. 
This is a natural consequence of the fact that the number of particles does
not depend on the observer.

To lowest order in $B$, the result in Eq.~\eqref{conversionprobability} coincides with the conversion probability in the
literature for which the BGMF-induced curvature is neglected, for example as in Ref.~\cite{Dolgov:2012be}.

\begin{figure}

\begin{center}
\includegraphics[clip,width=13.0cm]{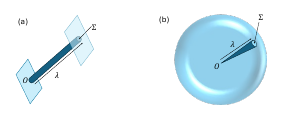}
\caption{(a) Congruence of null geodesics (dark cylinder)
    corresponding to a plane wave. The two plates perpendicular to
    the cylinder represent $\Sigma$ with different values of the phase.
    (b) Congruence of null geodesics (dark thin cone)
    corresponding to a spherical wave. The sphere perpendicular to
    the cone represents the surface with a fixed value of the phase.}
\label{fig:beams-configuration}
\end{center}
\end{figure}

\subsection{Case 1: plane waves }
\label{Case 1}
We first consider the case in which the configuration of the beam corresponds to a plane wave, as 
sketched in Fig.~\ref{fig:beams-configuration}(a).
A plane wave is the simplest and most fundamental solution of a wave equation. 
Thus, starting the analysis from this case provides a useful foundation for understanding the Gertsenshtein effect on the spacetime curved by BGMF. 
Furthermore, the analysis of the plane wave case is applicable to situations where a spherical wave emanating from a point source is locally approximated as a plane wave at sufficiently large distances, for which the expansion  $\Theta$ can be neglected.

We begin by computing the expansion scalar $\Theta$ up to $\mathcal{O}(B^2)$. 
We expand $\Theta$ in powers of the BGMF as 
$\Theta=\Theta_{(0)}+\Theta_{(2)}+\mathcal{O}(B^3)$. Here, the index $(i)$ denotes the term of $\mathcal{O}(B^i)$ for the indexed physical quantity. 
Note that quantities labeled with 
(0) refer to those evaluated in Minkowski spacetime.
$\hat{\sigma}_{\mu\nu}$ is also expanded in the same manner.

Assuming a spatially uniform and parallel beam at the origin, the initial conditions for the expansion and the shear are given by $\Theta(\lambda=0) = \hat{\sigma}_{\mu\nu}(\lambda=0) = 0$.
In Minkowski spacetime, the solution of Eqs.~(\ref{Expansion}) and (\ref{Shear}) with the above initial conditions is given by
\begin{equation}
\Theta_{(0)} (\lambda)=\hat{\sigma}_{(0) \mu\nu}(\lambda)=0.
\end{equation}
Since $\hat{\sigma}_{(0) \mu\nu}(\lambda)=0$,
$\hat{\sigma}_{ \mu\nu}(\lambda)$ starts at ${\cal O}(B^2)$ and 
the shear does not contribute to the evolution of $\Theta$ at ${\cal O}(B^2)$.
Thus, the equation for $\Theta_{(2)}$ becomes
\begin{align}
\frac{d}{d\lambda} \Theta_{(2)} &= - R_{\mu\nu} k^{\mu}_{(0)} k^{\nu}_{(0)} \nonumber \\
\label{Eq.Expansion(2)}&=-6\mathcal{R} \sin^2 \theta,
\end{align}
where $\mathcal{R}\equiv \frac{4\pi G}{3}B^2$ has been introduced again.
Solution of this equation with the initial condition $\Theta_{(2)}(\lambda=0)=0$
is given by
\begin{equation}
\Theta_{(2)}(\lambda)=-6\mathcal{R} \lambda\sin^2 \theta .
\end{equation}
Thus, $\Theta$ up to second order in $B$ is given by
\begin{equation}
\Theta (\lambda)=-6\mathcal{R}\lambda\sin^2 \theta .
\end{equation}
Note that the expansion in this case is independent of the
parameters $\alpha, \beta_1, \beta_2$ appearing in $g^{(B)}_{\mu \nu}$.
Plugging this expression into Eq.~(\ref{AMPA_expanded}) yields
\begin{align}
\mathcal{A}^2(\lambda)&=\mathcal{A}_0^2 
\exp \left( 1+3\mathcal{R}\lambda^2\sin^2 \theta\right)
\cos^2\left( \sqrt{3\mathcal{R}}\lambda\sin\theta \right) \nonumber
\\&=A_0^2 (1+3\mathcal{R}\lambda^2\sin^2\theta)(1-3\mathcal{R}\lambda^2\sin^2\theta )
\nonumber \\
&=A_0^2.
\end{align}

Here, from the first to the second line, both the ``exponential term'', 
arising from the deformation of the congruence, and the``cos term'', 
originating from the Gertsenshtein effect, 
have been expanded up to $\mathcal{O}(B^2)$, respectively.

As seen from the above computation,
the curvature of spacetime enhances the wave amplitude for any value of $\theta$ by focusing 
the congruence.
This feature is consistent with the fact that the electromagnetic field satisfies the strong energy condition.
On the other hand, the Gertsenshtein effect reduces the wave amplitude
for any value of $\theta$,
which is also expected because in the current setup no photons are
converted from gravitons by the Gertsenshtein effect.
Remarkably, these two physically independent effects exactly cancel each other, 
independently of the value of $\theta$. 
As a result, the amplitude of EMWs remains unchanged through propagation.
At present, we do not have a simple explanation for why this exact cancellation occurs. 

Before moving to the second example, we provide a consequence of our result which may be
relevant to experiments aiming at measuring or constraining magnetic fields via
the Gertsenshtein effect.
Consider a source emitting EMWs that can be approximately treated as a plane wave, 
and a distant observer who measures the EMWs.
In the literature, BGMF-induced curvature was neglected
in the computations of the propagation of waves in the BGMF,
for which the energy flux of the EMWs is reduced only by the Gertsenshtein effect.
Given that the amount of reduction depends on the strength of the magnetic field,
by comparing the observed energy flux  {($\mathcal{A}^2\omega^2_{\text{obs}}$)}  with the intrinsic (a priori known) flux {($\mathcal{A}^2_0\omega^2_0$)}, 
one could, in principle, estimate the strength of the magnetic field \footnote{This idea 
is implemented in Ref.~\cite{chen2013primordial}, where constraints on the cosmic magnetic field strength are obtained by estimating deviations of the CMB spectral distortion induced by the Gertsenshtein effect.}.
However, our result indicates that the BGMF-induced curvature 
affects the wave amplitude  to the same extent as the Gertsenshtein effect, and even completely 
cancels the effect of the photon-graviton conversion when the EMWs are perfectly plane wave.
Thus, in experiments that measure the flux of the EMWs,
the focusing of the congruence due to the spacetime curvature must be accounted for 
when interpreting the experimental data.
On the other hand, if the experimental setup is sufficiently large to capture the entire
region traversed by the EMWs and to measure the total energy (i.e., 
energy flux times the area), then the effect of spacetime curvature becomes irrelevant,
and the only physical effects are the energy loss due to the Gertsenshtein effect and the deviation of the observed frequency from $\omega_0$.
In such a case, one can in principle measure the magnetic field strength from the observed energy deficit.

\subsection{Case 2: spherical waves}
\label{Case 2}
The second example is the case in which a spherical electromagnetic wave is emitted from a
point-like source inside the magnetic field region, as 
sketched in Fig.~\ref{fig:beams-configuration}(b). 
Assuming isotropic emission, the initial conditions for
the expansion and the shear are given by
$\lim_{\lambda \to 0} \lambda \Theta (\lambda)=2, ~\hat{\sigma}_{\mu\nu}(\lambda=0)=0$. 
In Minkowski spacetime, the solution of Eqs.~(\ref{Expansion})
and (\ref{Shear}) with the above initial conditions is given by
\begin{equation}
\Theta_{(0)} (\lambda)=\frac{2}{\lambda},~~~~~~~\hat{\sigma}_{(0) \mu\nu}
(\lambda)=0.
\end{equation}
Since $\hat{\sigma}_{(0) \mu\nu}(\lambda)=0$,
$\hat{\sigma}_{ \mu\nu}(\lambda)$ starts at ${\cal O}(B^2)$ and 
the shear does not contribute to the evolution of $\Theta$ at ${\cal O}(B^2)$.
Thus, the equation for $\Theta_{(2)}$ becomes
\begin{align}
\frac{D}{D\lambda} \Theta_{(2)} &=
-\Theta_{(0)} \Theta_{(2)} - R_{\mu\nu} k^{\mu}_{(0)} k^{\nu}_{(0)} \nonumber \\
&=-\frac{2}{\lambda} \Theta_{(2)}-6\mathcal{R} \sin^2 \theta.
\end{align}
Solution of this equation with the initial condition $\Theta_{(2)}(\lambda=0)=0$
is given by
\begin{equation}
\Theta_{(2)}(\lambda)=-2\mathcal{R} \sin^2 \theta \lambda.
\end{equation}
Thus, $\Theta$ up to second order in $B$ is given by
\begin{equation}
\Theta (\lambda)=\frac{2}{\lambda}(1-\mathcal{R}\lambda^2 \sin^2 \theta )
.
\end{equation}
Note that the expansion in this case is also independent of the
parameters $\alpha, \beta_1, \beta_2$ appearing in $g^{(B)}_{\mu \nu}$.
Plugging this expression into Eq.~(\ref{AMPA_expanded}) yields
\begin{align}
\mathcal{A}^2(\lambda)&=\mathcal{A}_0^2 
\exp \left( -2\ln \frac{\lambda}{\lambda_0}+\mathcal{R}(\lambda^2-\lambda_0^2)\sin^2 \theta\right)
\cos^2\left( \sqrt{3\mathcal{R} }\lambda\sin\theta \right) \nonumber
\\&=
{A_0^2\frac{\lambda_0^2}{\lambda^2} (1+\mathcal{R}\lambda^2\sin^2\theta)(1-3\mathcal{R}\lambda^2 \sin^2\theta )}
\nonumber \\
&= {\mathcal{A}^2_0 \frac{\lambda_0^2}{\lambda^2} (1-2\mathcal{R}\lambda^2\sin^2\theta)}
,
\end{align}
where $\lambda_0$ is {a positive infinitesimal} introduced
to regularize the divergent integral $\int_0^\lambda \Theta (\lambda')d\lambda'$
to $\int_{\lambda_0}^\lambda \Theta (\lambda')d\lambda'$.
In this regularization, $\mathcal{A}_0$ is the amplitude of $\mathcal{A}$
at $\lambda=\lambda_0$.
Because $\mathcal{A} (\lambda) \propto 1/\lambda$ in Minkowski spacetime,
$\mathcal{A}_0 \lambda_0$ is independent of $\lambda_0$ for sufficiently
small $\lambda_0$, guaranteeing $\mathcal{A}(\lambda)$ given by the
above equation is independent of the choice of $\lambda_0$.
Contrary to the case of a plane wave, $\mathcal{O}(B^2)$ term remains 
non-vanishing in $\mathcal{A}^2$.
More specifically, the focusing effect caused by the curvature of spacetime is only one-third of the Gertsenshtein effect at any angle $\theta$,
and the net effect is a reduction in the magnitude of $\mathcal{A}^2$.

\section{Conclusions}
In this work, we studied the Gertsenshtein effect on spacetime curved by a background magnetic field. Under the geometric optics approximation, we derived the exact equations describing the evolution of EMWs and GWs during the propagation, valid up to order $\mathcal{O}(B^2)$.

In Sec.~II, we determined the background metric induced by a static and uniform magnetic field along the $z$-axis by solving the linearized Einstein equations. This cylindrically symmetric metric contains three free parameters, $\alpha$, $\beta_1$, and $\beta_2$: $\beta_1$ and $\beta_2$ correspond to the gauge degrees of freedom, while $\alpha$ is fixed by the boundary conditions.

In Sec.~III, we derived the evolution equations for the polarization vector/tensor and the amplitude of EMWs and GWs from the linearized Einstein--Maxwell equations, respectively. The polarization vector/tensor is shown to evolve via parallel transport along null geodesics. The amplitude equations can be decomposed into two independent polarization states, each satisfying the same form of the evolution equations. The amplitude evolution is influenced by two effects: the convergence/divergence of geodesics and the conversion between EMWs and GWs. Furthermore, the total number of photons plus gravitons is conserved throughout the propagation.

In Sec.~IV, we analyzed the evolution of the energy flux of EMWs by solving the amplitude equations derived in Sec.~III, assuming an initial state with only EMWs. The amplitude squared of EMWs, proportional to the energy flux, was computed in two specific cases: plane waves and spherical waves. The evolution is governed by the net effect of the two contributions mentioned in Sec.~III. In the plane wave case, the amplitude squared of EMWs remains constant during propagation, regardless of the angle between the BGMF and the direction of the injected EMWs. 
On the other hand, in the spherical wave case, the EMW amplitude squared decreases. This result arises because the magnification effect from the deformation of the congruence is weaker than the attenuation effect caused by the conversion of EMWs into GWs.
These results for the two specific cases are independent of the free parameters of the background metric. The independence from the gauge parameters is a general conclusion, because gauge degrees of freedom do not affect the physical dynamics. On the other hand, the independence from the parameter $\alpha$ is specific to these two cases.
It must finally be noted that while the wave amplitude is influenced by the curvature of spacetime
induced by the background magnetic field, we found that the conversion probability,
defined in terms of particle number, remains unaffected.

\section*{Acknowledgments}
This work was supported by JST SPRING, Japan Grant Number JPMJSP2180. P.G. was supported in part by the National Science Foundation award PHY-2412829, and acknowledges the hospitality and support of the Institute of Science Tokyo.
T.S. gratefully acknowledges support from JSPS KAKENHI grant (Grant Number JP23K03411)

\appendix
\section{Visualization of the cross-sectional area}
\label{Appendix:visualCS}
In the main text, we adopted a description based on the quantities 
$\{\Theta, \sigma_{\mu\nu}\}$ to characterize the beam. 
This approach offers a simple and straightforward method for computing the evolution of the expansion scalar $\Theta$.
In this Appendix, in order to visually understand the deformation of the congruence, we employ an alternative formulation.

To begin with, we construct the geodesic congruence and the cross-sectional area in terms of deviation vectors. We introduce the parameters $(s,t)$ to label the geodesics in the congruence, where the central geodesic $x^\mu(\lambda)$ is specified by $(s=0,t=0)$, and coincides with the central geodesic in Sect.~\ref{def:bases}. In this notation, an arbitrary point on the congruence is specified by $x^\mu(\lambda,s,t)$.

The deviation vectors are defined as
\begin{equation}
    \xi^\mu_u = \partial_u x^\mu,
\end{equation}
where \(u\) denotes either \(s\) or \(t\). The vectors $\xi^\mu_s \, ds$ and $\xi^\mu_t \, dt$ describe how neighboring geodesics labeled \((s=0,t=0)\) and \((ds,0)\) or \((0,dt)\) deviate from each other, where $ds,dt$ are infinitesimal parameters. Accordingly, $\xi^\mu_u$ encode the local deformation of the congruence.

The cross-sectional area of the congruence is defined by the two-dimensional surface spanned by the deviation vectors \(\xi^\mu_s\) and \(\xi^\mu_t\). To visualize this area on the hypersurface \(\Sigma\), we expand the deviation vectors in terms of the orthonormal basis \(\{e_\mu^1, e_\mu^2\}\) introduced in Sec.~\ref{def:bases}:
\begin{equation}
    \xi^\mu_u = \xi_u^{a} e_a^\mu,
\end{equation}
where $(a = 1, 2)$. Since this basis is normalized, the evolution of the projected components \(\xi_u^{a}\) directly reflects the physical size and shape of the cross-sectional area.

The evolution of the projected components \(\xi_u^{a}\) is governed by the geodesic deviation equation:
\begin{align}
\label{Eq.deviation} \frac{D^2}{D\lambda^2}\xi^\mu=R^\mu{}_{\nu\rho\lambda}k^\nu k^\rho \xi^\lambda.
\end{align}
From the above equation, the projected components $\xi_u^a$ up to order $\mathcal{O}(B^2)$ are obtained as follows:
\begin{align}
\frac{d^2}{d\lambda^2} \xi^a_u
&= R^\mu{}_{\nu\rho\sigma} \, e_{(0)\mu}^{a} \, k_{(0)}^\nu \, k_{(0)}^\rho \, e^\sigma_{(0)b} \, \xi^b_u \,.
\end{align}
Substituting the explicit expressions of $R^\mu{}_{\nu\rho\sigma} , e_{(0)\mu}^{a} , k_{(0)}^\mu$, the explicit form of the equation of the projected components becomes
\begin{align}
\label{Evo:deviation}\frac{d^2 }{d\lambda^2} \begin{bmatrix}
    \xi_u^\theta \\
    \xi_u^\varphi
\end{bmatrix}
=
\begin{bmatrix}
-3 (1+\alpha) \mathcal{R} \sin^2\theta & 0 \\
 0 & -3 (1-\alpha) \mathcal{R} \sin^2\theta
\end{bmatrix}
\begin{bmatrix}
    \xi_u^\theta \\
    \xi_u^\varphi
\end{bmatrix}.
\end{align}
This equation is independent of the parameters corresponding to the gauge parameters \(\beta_1, \beta_2\), because only the background Riemann tensor contributes in Eq.~\eqref{Eq.deviation}. This ensures that the physical size of the cross-sectional area is gauge invariant.

In this way, the congruence is fully specified by the initial conditions on the deviation vectors together with the restrictions on the parameters $(s,t)$. The physical cross-sectional area is then obtained by projecting $\xi_u^a$ onto the plane spanned by $\{e_\mu^1,e_\mu^2\}$.

In the following subsection, we examine the evolution of the cross-sectional area. First, we construct two specific beam configurations, as described in Sec.~\ref{Sec.solution}, by specifying the initial conditions of the deviation vectors. We then compute the evolution of the projected components $\xi_u^a$ using Eq.~\eqref{Evo:deviation}, and finally visualize the cross-sectional area on the plane $\Sigma$ to analyze the cross-section deformation during propagation.

\subsection{Plane waves}
Assuming a spatially uniform and parallel beam at the origin, as we mentioned in Sec.~\ref{Case 1}, the initial conditions for the projected components \(\xi_u^a\) are given by
\begin{align}
    \xi_s^a = \qty{1, 0}, \quad \xi_t^a = \qty{0, 1}, \quad
\frac{d \xi_s^a}{d\lambda} = \qty{0,0}, \quad \frac{d \xi_t^a}{d\lambda} = \qty{0,0}.
\end{align}
Solving Eq.~\eqref{Evo:deviation} under the above initial conditions, we obtain the evolution of the projected components \(\xi_u^a\) as follows:
\begin{align}
    \xi_s^a =\qty{ 1-\frac{3(1+\alpha)}{2}  \mathcal{R}\lambda^2\sin^2\theta,0 },\quad \xi_t^a =\qty{0,1-\frac{3(1-\alpha)}{2} \mathcal{R}\lambda^2 \sin^2\theta}.
    \label{eq:xi-planar}
\end{align}
The vector \(\xi_s^\mu\) has components only along the \(e^1_\mu\) direction, while \(\xi_t^\mu\) has components only along the \(e^2_\mu\) direction. This is because the matrix acting on the vector \((\xi_u^{a=1}, \xi_u^{a=2})^\mathrm{T}\) in Eq.~\eqref{Evo:deviation} is diagonal.

\begin{figure}[htbp]
  \centering
\includegraphics[width=1\linewidth]{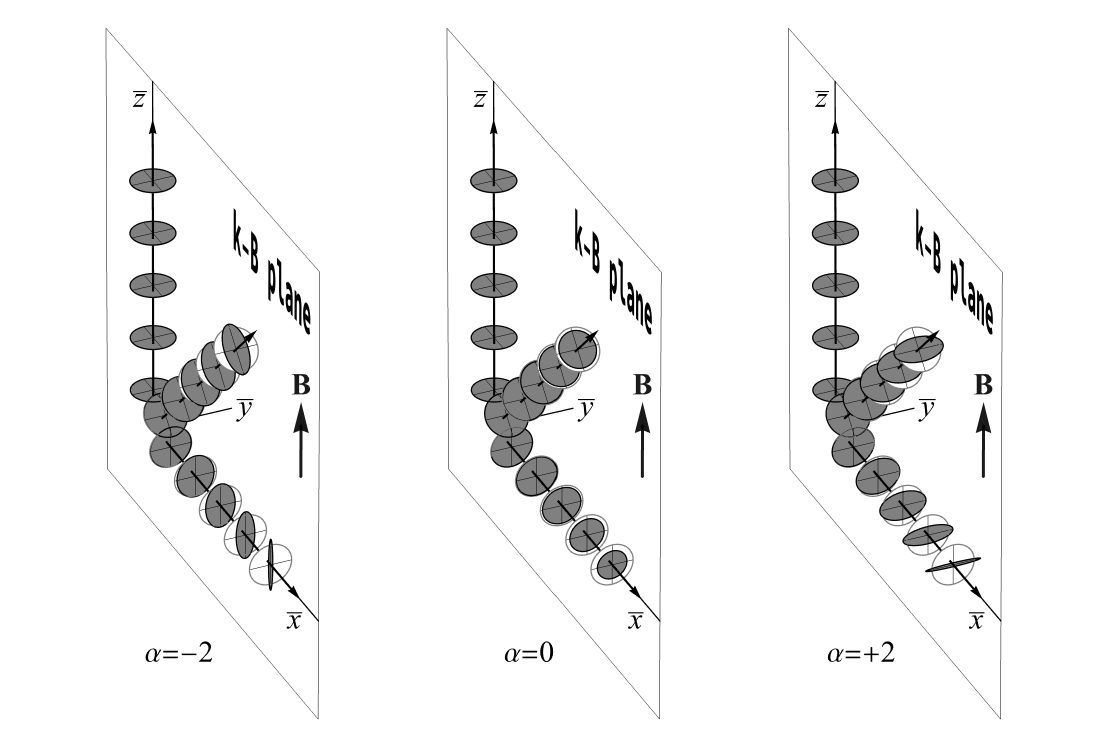}
 \caption{
This figure illustrates the evolution of the cross-sectional area during the propagation of a null congruence constructed from geodesics parameterized by \((s, t)\) satisfying \(s^2 + t^2 < 1 \) 
on the Riemann normal coordinate system
(see also Appendix \ref{appendix:tetrad}). 
Each panel (left to right) corresponds to different values of the free parameter 
$\alpha$ in $g^{(B)}_{\mu\nu}$. The chosen values
$\alpha = -2, 0, 2$ represent the behavior of the cross-sections in the regions
$\alpha < -1$, $-1 < \alpha < 1$, and $\alpha > 1$, respectively.
Geodesics through the origin in directions \(\theta = 0, \pi/4, \pi/2\) are shown as black straight lines on the \(\bar{x}\)-\(\bar{z}\) plane (i.e., the $\vec{k}$--$\vec{B}$ plane).
Along each geodesic, the cross-sections are illustrated by filled-in ellipses centered at $\overline{x}^\mu(\lambda)$ with $80 \lambda = 12,29,46,63,80$ (the cross-sections at $\lambda=0$ are omitted to avoid confusing overlapping regions). The shape of each cross section is to be judged against a circle of radius 1 in the plane of the cross section (the white disk accompanying each filled-in ellipse). 
}
\label{CSplane}
\end{figure}
Here, the null congruence is constructed by selecting geodesics labeled by parameters \((s, t)\) restricted to the region \(s^2 + t^2 < 1\). Figure~\ref{CSplane} visualizes the evolution of this cross-sectional area on Riemann normal coordinate system $(\bar{t},\bar{x},\bar{y},\bar{z})$ centered at the origin of $x^\mu(\lambda)$ \footnote{In these Riemann normal coordinates, the center geodesic $x^\mu(\lambda)$ through the origin is always a straight line, and the $e^2_\mu$ component has only $\bar{t}$ and $\bar{y}$ components. The explicit expressions for the components are provided in the Appendix~\ref{appendix:tetrad}.}.

The shape of the evolving cross-sectional area can be classified into three characteristic types, depending on the free parameter \(\alpha\) determined by the boundary conditions discussed in Sec.~\ref{Sec:metric}:

\begin{itemize}
    \item For \(\alpha < -1\) (left panel), the cross-sectional area is stretched along the \(e^1_\mu\) (horizontal) direction.
    \item For \(-1 < \alpha < 1\) (center panel), it is contracted along both \(e^1_\mu\) and \(e^2_\mu\)  direction.
    \item For \(\alpha > 1\) (right panel), it is stretched along the \(e^2_\mu\) (vertical) direction.
\end{itemize}

For each panel, the deformation depends on the angle \(\theta\) between the geodesic and the \(z\)-axis. At \(\theta = 0\), no deformation occurs; the second-order corrections in \(B\) do not affect the projected component \(\xi^a_u\), and the behavior is identical to that in Minkowski spacetime. As \(\theta\) increases, the magnitude of the deformation grows for the same affine parameter \(\lambda\). This effect reaches its maximum when the propagation direction of the beam is orthogonal to the background magnetic field, i.e., at \(\theta = \pi/2\).

Importantly, the physical cross-sectional area, given by
\begin{align}
    S &\equiv \pi \xi_s^1 \xi_t^2=\pi (1-3\mathcal{R}\lambda^2\sin^2\theta)
\end{align}
is independent of \(\alpha\). This follows from the fact that the expansion scalar does not depend on \(\alpha\), as shown in Sec.~\ref{Sec:metric}.

Therefore, \(\alpha\) affects only the shape of the congruence via the shear up to order \(\mathcal{O}(B^2)\), without altering the physical size of the cross-sectional area.

\subsection{Spherical waves}
\begin{figure}
    \centering
  \includegraphics[width=1\linewidth]{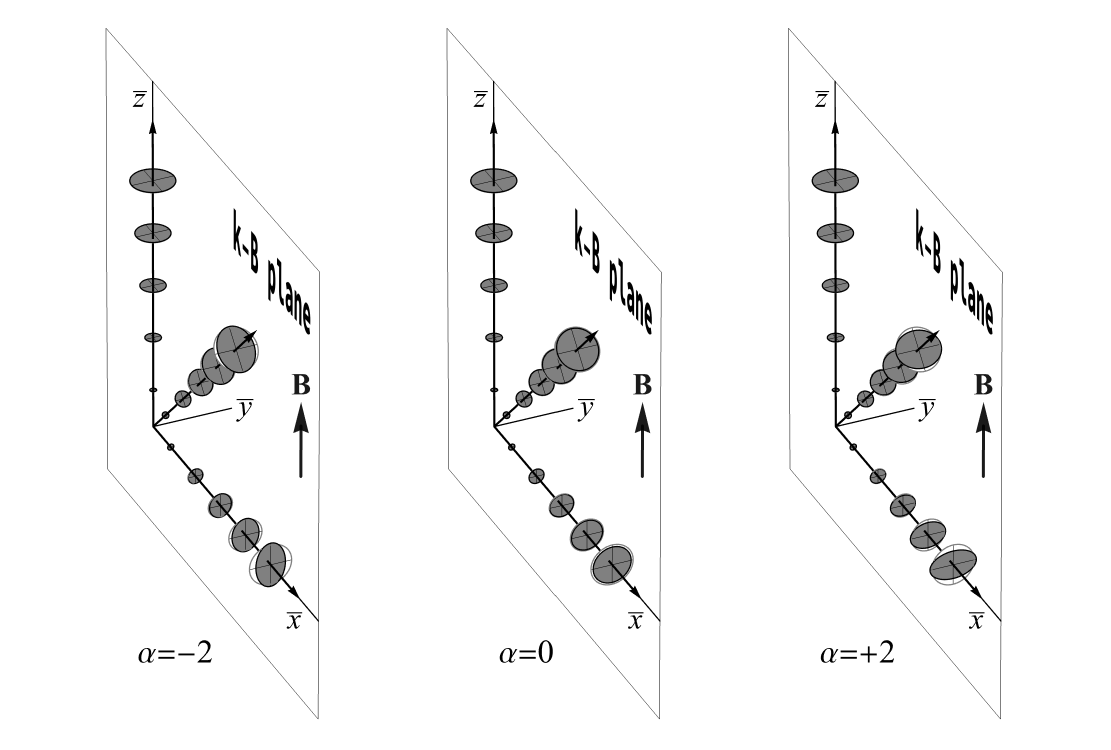}
    \caption{As in Fig.~\protect\ref{CSplane}, but for spherical waves emanating from the origin and the radii of the reference circles rescaled by $\lambda$ to better follow the spherical expansion of the waves.}
    \label{CSspherical}
\end{figure}

Assuming isotropic emission from a source at the origin, the initial conditions for the projected components \(\xi_u^a\) are given by
\begin{align}
    \xi_s^a = \qty{0, 0}, \quad \xi_t^a = \qty{0, 0}, \quad
\frac{d \xi^s_a}{d\lambda} = \qty{1,0}, \quad \frac{d \xi_t^a}{d\lambda} = \qty{0,1}.
\label{eq:xi-spherical}
\end{align}
Solving Eq.~\eqref{Evo:deviation}, we obtain the evolution of $\xi^a_s,\xi^a_t$
\begin{align}
    \xi_s^a =\qty{\lambda-\frac{(1+\alpha)}{2}\mathcal{R}\lambda^3\sin^2\theta,0 },\quad \xi_t^a =\qty{0,\lambda-\frac{(1-\alpha)}{2}\mathcal{R}\lambda^3\sin^2\theta}.
\end{align}
Here, the null congruence is constructed by selecting geodesics labeled by parameters \((s, t)\) restricted to the region \(s^2 + t^2 < 1\). Figure~\ref{CSspherical} visualizes the evolution of this cross-sectional area.

Unlike the case of plane waves, the cross-sectional area evolves isotropically with the affine parameter \(\lambda\) at the leading order. This behavior is observed along the geodesic at \(\theta = 0\); the cross-sectional area of the beam emitted in this direction is not affected by second-order corrections in \(B\), similar to the plane wave case. Moreover, for the same \(\lambda\), taking the circle at \(\theta = 0\) as a reference, the cross-sectional area exhibits anisotropic deformations depending on the value of \(\alpha\), as discussed in the plane wave case. This effect increases with \(\theta\) and reaches its maximum when the magnetic field and the direction of the beam are orthogonal.

The physical cross-sectional area is given by
\begin{align}
    S &\equiv \pi  \xi_s^1 \xi_t^2=\pi \omega_0^2\lambda^2(1-\mathcal{R}\lambda^2\sin^2\theta).
\end{align}
This is also independent of $\alpha,\beta_1,\beta_2$ appearing in $g^{(B)}_{\mu\nu}$ for the same reason discussed in the plane wave case.

\section{Christoffel symbols and the Riemann tensor}\label{Appendix:R}
In this appendix, we list the components of the Christoffel symbols $\Gamma^\mu_{\nu\lambda}$ and of the Riemann tensor $R_{\mu\nu\rho\sigma}$ corresponding to the metric defined in Sec.~\ref{Sec:metric}. The explicit expressions for the Christoffel symbols and the Riemann tensor are not required for the discussion in the main text; however, the explicit form of the Christoffel symbols is necessary for solving the equation of parallel transport in the Appendix.~\ref{appendix:tetrad}. We show the non-zero components of Christoffel symbols, omitting components that can be obtained by index symmetries from those listed.

\begin{align*}
\Gamma^t_{t\rho} &= -\alpha \mathcal{R} \rho,  
& \Gamma^t_{tz} &= (2\alpha+3)\mathcal{R} z, 
& \Gamma^\rho_{tt} &= -\alpha \mathcal{R} \rho, 
& \Gamma^\rho_{\rho\rho} &= \tfrac{1}{2}(2\alpha-3)\mathcal{R}\rho, 
\\[0.5em]
\Gamma^\rho_{\varphi\varphi} &= -\tfrac{1}{2}(2\alpha-3)\mathcal{R}\rho^3 - \rho, 
& \Gamma^\rho_{\rho z} &= -\mathcal{R} z(\alpha+\beta_1), 
& \Gamma^\rho_{zz} &= -\beta_1 \mathcal{R}\rho, 
& \Gamma^\varphi_{\rho\varphi} &= \tfrac{1}{2}(2\alpha-3)\mathcal{R}\rho + \tfrac{1}{\rho}, 
\\[0.5em]
\Gamma^\varphi_{\varphi z} &= -\mathcal{R} z(\alpha+\beta_1), 
& \Gamma^z_{tt} &= (2\alpha+3)\mathcal{R} z, 
& \Gamma^z_{\rho\rho} &= \mathcal{R} z(\alpha+\beta_1), 
& \Gamma^z_{\rho z} &= \beta_1 \mathcal{R}\rho, 
\\[0.5em]
\Gamma^z_{\varphi\varphi} &= \mathcal{R}\rho^2 z(\alpha+\beta_1), 
& \Gamma^z_{zz} &= \beta_2 \mathcal{R} z. 
& &
\end{align*}

The explicit form of the Riemann tensor is needed to derive Eq.~\eqref{Evo:deviation} in the Appendix.~\ref{Appendix:visualCS}.
We show the non-zero components of Riemann tensor, omitting components that can be obtained by index symmetries from those listed.
\begin{align*}
R_{t\rho t\rho}         &= -\mathcal{R} \alpha,           & 
R_{t\varphi t\varphi}   &= -\mathcal{R} \alpha \rho^2,    & 
R_{tz tz}               &= \mathcal{R}(3 + 2\alpha),      \\
R_{\rho\varphi \rho\varphi} &= \mathcal{R}(3 - 2\alpha) \rho^2, &
R_{\rho z \rho z}       &= \mathcal{R} \alpha,            & 
R_{\varphi z \varphi z} &= \mathcal{R}\alpha \rho^2.
\end{align*}
These expressions show that the Riemann tensor is independent of \( \beta_1, \beta_2 \), which reflects that gauge freedom does not affect the curvature.

\section{Central geodesic $x^\mu(\lambda)$, base vectors $e^1_\mu,e^2_\mu$, and Riemann normal coordinates}
\label{appendix:tetrad}
In the analysis in the main part of the paper, the leading-order expressions of \(k^\mu\), \(e^1\), and \(e^2\) up to \(\mathcal{O}(B^0)\) are sufficient. Nevertheless, for completeness and future reference, we provide their explicit expressions up to \(\mathcal{O}(B^2)\). These expressions are obtained by perturbatively solving the equation of parallel transport.

We impose the initial conditions  for the geodesic at $\lambda=0$ as stated in the main text: $x^\mu(0) = 0, k^\mu(0) = k^\mu_{(0)}(0).$
In Minkowski spacetime, the geodesic tangent vector remains constant: \( k^\mu_{(0)}(\lambda) = k^\mu_{(0)}(0) \).
To compute the second-order correction \( k^\mu_{(2)} \), we solve the geodesic equation \eqref{GWs-geodesic-eq} perturbatively in the affine parameter $\lambda$:
\begin{equation}
    \frac{d k^\mu_{(2)}}{d \lambda} = - \Gamma^\mu_{\nu \lambda} \, k^\nu_{(0)} \, k^\lambda_{(0)}.
\end{equation}
Integrating this equation with the given initial conditions yields \( x^\mu(\lambda) \) up to \( \mathcal{O}(B^2) \). The resulting expression is
\begin{align}
x^\mu(\lambda) = 
\begin{pmatrix}
\lambda\left[ 1-\frac{1}{3} \mathcal{R}\lambda^2\left( (3 + 3\alpha) \cos^2 \theta -\alpha \right) \right] \\[1ex]
\lambda\sin\theta  \left[ 1 + \frac{1}{4} \mathcal{R}\lambda^2 \left( 1 + (2\alpha + 2\beta_1 - 1) \cos^2 \theta \right) \right] \\[1ex]
0 \\[1ex]
\lambda\cos\theta  \left[ 1 -\frac{1}{6}\mathcal{R}\lambda^2 \left( 3 + 2\alpha + \beta_2 + (\alpha + 3\beta_1 - \beta_2) \sin^2 \theta \right) \right]
\end{pmatrix}.
\end{align}

Similarly, for the basis vectors \( e^a_\mu \), we impose the initial conditions at \( \lambda = 0 \) given in  Eq.~\eqref{Initial e}. In Minkowski spacetime, the the basis vectors \( e^a_\mu \) are constant: $ e^a_{(0)}{}_\mu(\lambda)=e^a_{(0)}{}_\mu(0)$ along the geodesic. Substituting into the parallel transport equation, we obtain the equation for $e^a_{(2)}{}_\mu$: 
\begin{equation}
    \frac{d e^a_{(2)}{}_\mu}{d \lambda} = \Gamma^\nu_{\mu \lambda} \, k^\lambda_{(0)} \, e^a_{(0)}{}_\nu,
\end{equation}
Solving this equation under the given initial conditions, the expression of $e^a_\mu$ is given by

\begin{align}
e^1_\mu(\lambda) & = 
\begin{pmatrix}
- \frac{3}{4} \mathcal{R} \lambda^2 (1 + \alpha)  \sin(2\theta) \\[1ex]
\cos\theta \left[ 1 - \frac{1}{4} \mathcal{R} \lambda^2 \left( 2\alpha + 2\beta_1 + (3 - 2\alpha + 2\beta_1) \sin^2 \theta \right) \right] \\[1ex]
0 \\[1ex]
- \sin\theta \left[ 1 + \frac{1}{2} \mathcal{R} \lambda^2 \left( \beta_1 + (\alpha + \beta_1 + \beta_2) \cos^2 \theta \right) \right]
\end{pmatrix}\\
e^2_\mu(\lambda) & = 
\begin{pmatrix}
0 \\[1ex]
0 \\[1ex]
1 - \frac{1}{4} \mathcal{R} \lambda^2 \left( 2 \alpha + 2 \beta_1 + (3 - 4 \alpha - 2 \beta_1) \sin^2 \theta \right) \\[1ex]
0
\end{pmatrix}
\end{align}
In the illustrations in Figs.~\ref{CSplane} and~\ref{CSspherical}, we use a Riemann normal coordinate system $\overline{x}^\mu$ centered at the origin of $x^\mu$. Writing the metric in Eq.~(\ref{sec2:cylindrical-metric-2}) in the compact form
\begin{align}
    g_{\mu\nu} = \eta_{\mu\nu} + g_{\mu\nu\alpha\beta} x^\alpha x^\beta ,
\end{align}
our coordinate transformation to Riemann normal coordinates reads
\begin{align}
    \overline{x}^\mu = x^\mu + \frac{1}{6} J^{\mu}_{\ (\alpha\beta\gamma)} x^\alpha x^\beta x^\gamma,
\end{align}
where
\begin{align}
    J_{\mu\alpha\beta\gamma} = g_{\mu\beta\gamma\alpha} + g_{\mu\gamma\beta\alpha} - g_{\beta\gamma\mu\alpha} .
\end{align}

In the  $\overline{x}^\mu$ coordinate system, a geodesic through the origin with initial tangent vector $k^\mu$ is a straight line with parametric equation
\begin{align}
    \overline{x}^\mu = \lambda k^\mu .
\end{align}
In particular, with our choice of $k^\mu$,
\begin{align}
    \overline{x}^\mu(\lambda) =
    \begin{pmatrix}
 \lambda \\
 \lambda \cos\theta \\
 0\\
 \lambda \sin\theta
 \end{pmatrix}
\end{align}

The $\overline{e}^a_{\mu}$ components of the base vectors ($a=1,2$) follow from $\overline{e}^a_{\mu} = e^a_{\nu} \, (\partial x^\nu/\partial \overline{x}^\mu)$. With a further electromagnetic gauge transformation $\overline{e}^a_{\mu} \to \overline{e}^a_\mu - \overline{e}^a_{0} k_\mu$ to eliminate the time component, we find
\begin{align}
\overline{e}^1_{\mu}(\lambda) & =
\begin{pmatrix}
 0 \\
 \cos\theta \Bigl[ 1 + \tfrac{1}{2}\mathcal{R}\lambda^2 (1+\alpha)\sin^2\theta \Bigr] \\
 0 \\
 - \sin\theta \Bigl[ 1 + \tfrac{1}{2}\mathcal{R}\lambda^2 (1+\alpha)\sin^2\theta \Bigr]
\end{pmatrix}
\\
\overline{e}^2_{\mu}(\lambda) & =
\begin{pmatrix}
 0 \\
 0 \\
 1 - \frac{1}{2} \mathcal{R}\lambda^2 (\alpha-1)\sin^2\theta\\
 0
 \end{pmatrix}
\end{align}

The metric in Riemann normal coordinates follows from $\overline{g}_{\sigma\tau} = g_{\mu\nu} \, (\partial x^\mu/\partial \overline{x}^\sigma) \, (\partial x^\nu/\partial \overline{x}^\tau)$ as
\begin{align}
    \overline{g}_{\mu\nu} = \eta_{\mu\nu} + \overline{g}_{\mu\nu\alpha\beta} x^\alpha x^\beta
\end{align}
with
\begin{align}
    \overline{g}_{\mu\nu\alpha\beta} = \frac{1}{6} (R_{\mu\alpha\nu\beta} + R_{\mu\beta\nu\alpha} ) ,
\end{align}
where $R_{\mu\alpha\nu\beta} = g_{\mu\beta\alpha\nu} + g_{\alpha\nu\mu\beta} - g_{\mu\nu\alpha\beta} - g_{\alpha\beta\mu\nu}$ are the components of the Riemann tensor to first order in $\mathcal{R}$.

Moreover, in Figs.~\ref{CSplane} and~\ref{CSspherical}, a cross section in the tangent space at $\overline{x}^\mu(\lambda)$ is shown as the region interior to the curve 
\begin{align}
    \xi^1_s(\lambda) \overline{e}_1^\mu(\lambda) \cos\vartheta +\xi^2_t(\lambda) \overline{e}_2^\mu(\lambda) \sin\vartheta
    \qquad (0\leq \vartheta \leq 2\pi).
\end{align}
Here the $\xi^a_u$ are given in Eqs.~(\ref{eq:xi-planar}) and~(\ref{eq:xi-spherical}) for the planar and spherical cases, respectively. The reference circles used to judge the shape of the cross sections are the regions interior to the curves
\begin{align}
    \overline{e}_1^\mu(\lambda) \cos\vartheta +
    \overline{e}_2^\mu(\lambda) \sin\vartheta \qquad (0\leq \vartheta \leq 2\pi).
\end{align}
As follows from the orthonormality of the vectors $\overline{e}_a^\mu(\lambda)$, these are circles of radius 1 in the tangent space at $\overline{x}^\mu(\lambda)$. In the spherical wave case, each of these reference circles ahs been multiplied by $\lambda$ to better convey the spherical expansion of the waves.

\bibliographystyle{unsrt} 
\bibliography{references}

\end{document}